\documentclass[superscriptaddress,groupedaddress,nofootnoteinbib,12pt]{article}  

\usepackage{amsmath}
\usepackage{amsfonts}
\usepackage{authblk}
\usepackage{bbm}
\usepackage{color}
\usepackage[dvipsnames]{xcolor}
\usepackage{graphicx,graphics}
\usepackage{epstopdf}
\usepackage{caption}
\usepackage{subcaption}
\usepackage{float} 
\usepackage{pifont}
\usepackage{titlesec}
\usepackage{etoolbox}
\usepackage{indentfirst}
\usepackage{jcappub}
\usepackage{epstopdf}
\usepackage{braket}

\def\bb{\begin{equation}}
\def\ee{\end{equation}}
\def\ba{\begin{array}}
\def\ea{\end{array}}
\def\babc{\begin{subequations}}
\def\eabc{\end{subequations}}

\def\5{\hspace*{5mm}}
\def\2{{\scriptstyle\frac12}}

\begin{document}

\begin{flushright}
NYU-TH-12/31/17
\end{flushright}

\vspace{0.5in}

\begin{center}

\Large{\bf Holographic CBK Relation}

\vspace{0.5in}

\large{Gregory Gabadadze and Giorgi Tukhashvili}

\vspace{0.2in}

\large{\it Center for Cosmology and Particle Physics, Department of Physics, \\
New York University, 726 Broadway,  New York, NY, 10003}

\vspace{0.3in}

\end{center}

The Crewther-Broadhurst-Kataev (CBK) relation connects the Bjorken function for deep-inelastic
sum rules (or the Gross - Llewellyn Smith function) with  the Adler function   for  electron-positron
annihilation in QCD; it has been checked to hold up to  four  loops in perturbation theory.
Here we study non-perturbative  terms in  the CBK relation using a holographic dual
theory that is believed to capture properties of QCD. We show that
for the large invariant momenta the perturbative CBK relation is  exactly satisfied.  For the small
momenta non-perturbative corrections enter the relation  and we calculate their significant
effects.   We also give an exact holographic  expression for the Bjorken function, as well as for the
entire three-point axial-vector-vector correlation function, and check their consistency
in the conformal limit.

\newpage



\section{Introduction and summary}


Our goal is to derive from holography a non-perturbative analogue of the Crewther-Broadhurst-Kataev (CBK) relation \cite{Crewther:1972kn,Broadhurst:1993ru}.  In perturbative QCD (pQCD) with an arbitrary number of colors and flavors
the CBK relation connects the  Bjorken function for deep-inelastic sum rules (or the Gross-Llewellyn Smith (GLS) function) with  the Adler function  for the electron-positron
annihilation.  While the two functions have perturbative corrections of their own,
most of those corrections cancel out  in their product; the  remaining ones assemble themselves
into the QCD $\beta$-function $\beta (\alpha_s)$, as  was
first discovered  by Broadhurst and Kataev in \cite {Broadhurst:1993ru} in the three-loop approximation.
Hence, the  CBK relation reads:
\bb\label{CBK}
C_{Bj} (\alpha_s) \cdot  D_5 (\alpha_s) =1 + \beta (\alpha_s) r (\alpha_s)\,.
\ee
Here $r$ is a polynomial of its argument, that also depends  on the
number of colors and flavors, and $C_{Bj}$ and $D_5$ are the Bjorken and Adler functions
defined in (\ref{vector_OPE}) and (\ref{axial_OPE}),
respectively.\footnote {Using the GLS function instead of $C_{Bj}$ would lead to
additional terms on the right hand side (RHS) of (\ref {CBK}) due to the light-by-light type diagrams; for simplicity
we will focus on the Bjorken function in what follows.}
It was subsequently argued that the CBK relation should be satisfied to all orders in pQCD
\cite {Gabadadze:1995ei,Crewther:1997ux,Braun:2003rp,Kataev:1996ce}, while the explicit calculations were done up to  four loops \cite{Baikov:2012zn}, confirming the validity of the relation to that order
(see also \cite{Brodsky:1995tb,Kataev:2010du,Cvetic:2016rot} for further work on various
aspects of the  CBK relation in pQCD).  However, the fate of non-perturbative
corrections to the CBK relation -- of potential importance  from the particle physics
pint of view \cite{Brodsky:1995tb} -- have not been studied.

In this paper we study the status of the CBK relation at the non-perturbative  level.
While such a quest is possible within QCD itself, in an order-by-order  Operator Product Expansion
(OPE) of the  correlators relevant for the Bjorken and Adler functions,
we will adopt a  different path  that enables exact calculations via the  holographic duality  \cite{Maldacena:1997re,Gubser:1998bc,Witten:1998zw}.  The latter  is a powerful framework to study strongly coupled  conformal  quantum field theories  with
the large number of colors $N_c$.  The real world QCD is neither  strongly coupled (at high energies) nor
is conformal, and has $N_c=3$. Nevertheless, it is useful to think of a conformal cousin of QCD,
with an arbitrary number of colors and flavors; in this case, the CBK relations should simplify
to, $ C_{Bj} (\alpha_s) D_5 (\alpha_s) =1$. As to the strong coupling imposed by holographic computations,
one  could only rely on an assumption that no singular behavior is encountered as one  interpolates from a
strongly coupled regime to a weakly coupled one.

There  exist two approaches to holographic QCD in the literature:
the top-down approach where a string theory
model generates symmetries and degrees of freedom relevant to QCD via intersecting
D-branes (see, e.g., \cite{Witten:1998zw,Sakai:2005yt} and  works referencing them),
or a  bottom-up approach in which the minimal
field content of the gravity dual is postulated  just to capture  the relevant physics of QCD
(see, \cite{Erlich:2005qh,DaRold:2005mxj,Hirn:2005nr}). Neither of the  above
two approaches is ideal:  the former typically generates other particles
at the scale of QCD, and hence  describes a  strongly coupled theory that is significantly
different from QCD, while the latter approach is  less justifiable {\it a priori}.

Nevertheless, both of the above approaches have been used,  and  often produced
reasonable agreement with the know QCD results. In particular, the bottom-up models were
successful in describing the spectrum of the light mesons and the correlation functions of
quark currents \cite{Erlich:2005qh,DaRold:2005mxj,Hirn:2005nr}. Since our quest in this work
has relevance to low energy hadron  physics, we adopt the bottom-up approach  (with all the
above-mentioned caveats in mind); in particular, we will use the model proposed in Ref. \cite{Hirn:2005nr}
amended by the 5D Chen-Simons term required to study the 4D correlation functions exhibiting the axial anomaly
\cite{Domokos:2007kt,Grigoryan:2007vg,Domokos:2009cq}.

We derive the CBK relation using the above holographic model.  The relation
explicitly exhibits the hadronic resonances in the vector and (non-siglet) axial channels,
and thus, captures non-perturbative dynamics of QCD. In the limit of the  large invariant momenta
we recover the perturbative CBK relation  of a conformal field  theory.
For the small invariant momenta non-perturbative corrections enter the relation
and give rise to the resonance behavior which is straightforwardly obtained. {\it En route}, we obtain
an exact holographic expression for the Bjorken function, as well as the
entire three-point axial-vector-vector correlation function.  The latter, we show, agrees in the
conformal limit with the one-loop  expression obtained in a conformal field theory.

\vspace{0.2in}


\section{A brief summary of the  pQCD results}


\vspace{0.1in}

Crewther \cite{Crewther:1972kn} found that two seemingly independent quantities, the Bjorken coefficient,
$C_{Bj}$, entering the deep-inelastic sum rules, and the Adler function  of the $e^+e^-$ annihilation, both
appearing in the OPE of quark currents, were  in fact related to each other. To make this explicit  we recall
the definitions of these two quantities  via respective OPE's:
\bb\label{vector_OPE}
i \int d^4 x e^{i p x}  T \left\{ J_\mu^{em} (x) J_\nu^{em} (0) \right\} =
\frac{2 i}{3} C_{Bj} \left( \ln \left( \frac{\mu^2}{-p^2} \right), \alpha_s (\mu^2) \right) \varepsilon_{\mu \nu \rho \sigma} \frac{p^\rho}{p^2}   J_5^{ (3) \sigma} (0) + \cdots
\ee
\bb\label{axial_OPE}
i \int d^4 x e^{i p x}  T \left\{ J_{5 \mu}^{(3)} (x) J_{5 \nu}^{(3)} (0) \right\} =
\left(  p^2 \eta_{\mu \nu} - p_\mu p_\nu \right) \frac{-N_c}{12 \pi^2} \frac{1}{2}
\Pi_5 \left( \ln \left( \frac{\mu^2}{-p^2} \right), \alpha_s (\mu^2) \right)  + \cdots\,,
\ee
where $J_\mu^{em}$ is a quark electromagnetic current,
$\alpha_s(\mu)$ is a scale dependent strong coupling constant,  and $N_c$ denotes
the number of colors. Furthermore, we ignore the electromagnetic loop corrections, and also the quark
masses; in this  approximation the flavor non-singlet axial current, $J_5^{ (3) \sigma}$, is conserved
and thus has zero anomalous dimension. Moreover, our $D_5$ then coincides, in the perturbative approximation,
with the Adler's $D$ function defined via the symmetric part of the vector-vector correlation function.
In the approximation of massless quarks and switched off electromagnetic loop corrections,
both $C_{Bj}$  and a log derivative of $\Pi_5$ (the analog of the Adler function, $D_5$) are renormalisation
group invariants and  satisfy
\bb
C_{Bj} \left( \ln\left( \frac{\mu^2}{-p^2} \right), \alpha_s (\mu^2) \right) = C_{Bj}
 \left(1, \alpha_s (P^2) \right)\,, \nonumber
 \ee
 \bb
{\partial \over \partial \, ln( -p^2) } \,\Pi_5 \left( \ln \left( \frac{\mu^2}{-p^2} \right), \alpha_s (\mu^2) \right) =
{\partial \over \partial  \,ln(P^2)}\,
\Pi_5 \left(1, \alpha_s (P^2) \right)\,\equiv D_5\,,
\ee
where $P^2=-p^2$, and $\alpha_s(P^2)$ denotes the QCD running coupling.
Now we sketch how the CBK relation emerges  in QCD (see \cite{Gabadadze:1995ei}, and
references therein).  One starts with  the anomalous three-point function and does
the  OPE of the two electromagnetic currents in the antisymmetric channel  with respect to their indices
(See Fig. \ref{OPE}):
\begin{figure}[t]
\centering
\includegraphics[width=0.5\textwidth]{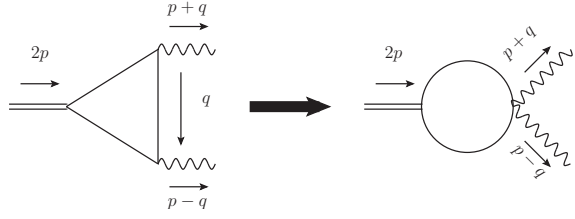}
\caption{A diagrammatic sketch of the OPE (\ref{three_point_OPE})}\label{OPE}
\end{figure}
\begin{align}\label{three_point_OPE}
{}  \Delta^{em,em,(3)}_{\mu \nu \rho} &= i \int d^4 x d^4 y e^{i k_1 x + i k_2 y}
\braket{0| T \left\{ J^{em}_\mu (x) J^{em}_\nu (y) J^{5  (3)}_\rho (0) \right\} | 0} \Big{|}_{q^2 \rightarrow + \infty} = \\
\nonumber = & \frac{N_c}{9 \pi^2 q^2} C_{Bj} \left(\alpha_s (Q^2) \right)
\Pi_5 \left(\alpha_s (4P^2) \right)
\left( p_\mu \varepsilon_{\nu \rho \sigma \tau} - p_\nu \varepsilon_{\mu \rho \sigma \tau} \right) p^\sigma q^\tau..\,,
\end{align}
here $Q^2=-q^2$ and we introduced new variables $k_1 = p +q$ and $k_2 = p-q$, with the  kinematical constraint $p \cdot q=0$ \cite{Gabadadze:1994tn}.  In the one loop approximation the three-point function  was calculated by Rosenberg \cite{Rosenberg:1962pp},  and its anomalous behavior was discovered  by Adler,  and by
Bell and Jackiw \cite{Adler:1969gk,Bell:1969ts}.\footnote{Schwinger's earlier work  \cite {Schwinger} derives an
expression for the axial anomaly via the background field method. However, the expression is obtained
as a  matrix element of the pseudoscalar operator, $2m i{\bar \psi} \gamma_5 \psi$, acting on two photon states
in the limit when all the momenta are negligible as compared to the physical fermion mass $m$. This coincides with
the axial anomaly (with an opposite sign) but this is not  the axial anomaly itself; the latter  should appear in addition to $2im {\bar \psi} \gamma_5 \psi$, and is missing in  Schwinger's work  \cite {Schwinger}. We  are grateful to Arkady Vainshtein for remembering and insisting that Schwinger had it wrong, contrary to the statement in our version 1.}

The Adler-Bardeen theorem on the one-loop nature of the axial anomaly \cite {Adler:1969er}, can be formulated as the existence  of a definition of the composite operator of the axial current for  which the anomaly in the operator form is exactly given by
the one loop diagram.\footnote{This definition however is not  unique and might or might not be enforced by
underlying symmetries in a full theory; a well-know example where it's not  enforced
is  in a supersymmetric theory.} Schreier \cite{Schreier:1971um} argued that if the theory is conformally invariant then one loop result is exact and no corrections from higher loops are allowed in  the entire three-point function (and not only in its anomalous part; conformal symmetry constrains  the form  of the three-point function up to a constant factor, which is fixed through the one-loop anomaly). In QCD this was studied at two loops \cite{Jegerlehner:2005fs} and  at three loops \cite{Mondejar:2012sz}. The two loop  result is null, while  Mondejar and Melnikov showed  that at  three loops there are nonzero terms proportional to the QCD beta function  \cite{Mondejar:2012sz}, consistent with an expectation from the CBK relation. Hence, gathering these results we can write:
\begin{align}
\nonumber {} & \frac{N_c}{9 \pi^2 q^2}
C_{Bj} \left( \alpha_s (Q^2) \right)
\Pi_5 \left(\alpha_s (4P^2) \right)= \\
\nonumber {} & =
 F_2^{PT} (p^2,q^2) \Big{|}_{q^2 \rightarrow + \infty}  + \frac{N_c}{9 \pi^2 q^2} \beta_{QCD} (\alpha_s (Q^2))
R \left( \alpha_s(4P^2), \,  \alpha_s(Q^2)\,  \right)\,,
\end{align}
where $F_2^{PT}$ is a  form factor within the anomalous three point function defined explicitly in
Appendix \ref{app_A},  while the function $R$ is some polynomial of its arguments.\footnote{
Strictly speaking  one could in general get  some  function of the beta-function
on the RHS, as long as that function vanishes when the beta-function does. For simplicity,
and also due to the three loop result \cite{Mondejar:2012sz},  we keep just the  beta-function.}
Taking now the logarithmic derivative with respect to $P^2$ we obtain:
\bb
C_{Bj} \left( \alpha_s (Q^2) \right) D_5 \left( \alpha_s (4P^2) \right) =
1 +  \beta_{QCD} (\alpha_s)  \tilde{R} \left( \alpha_s (4P^2),\, \alpha_s(Q^2) \right) \,,
\ee
with $\tilde R$ related to $R$. Furthermore, using  the definition
\bb
\alpha_s (4P^2) = \alpha_s (Q^2) - \int_{\ln {Q^2}}^{\ln{4P^2}} \beta_{QCD}(\alpha_s(e^u))du\,,
\ee
we  can schematically write:
\bb\label{gen_CBK}
C_{Bj} \left( \alpha_s (Q^2) \right) D_5 \left(  \alpha_s (Q^2) \right) = 1 +  \sum_{n=0} \sum_{j=1} \beta_n^j ~
r_{n,j} \left( \alpha_s (4P^2),\,\alpha_s(Q^2)\,, \ln \left( \frac{4 P^2}{Q^2} \right) \right)\,,
\ee
where $\beta's$ are the coefficients of the QCD beta function, and $r_{n,j}$'s some functions of their arguments.
Thus, in   a conformal theory the RHS equals to 1; this is the result we'd like to obtain from holographic QCD.

In what follows, we will also need the antisymmetric part of the two-point correlation function of the electromagnetic currents. If we just sandwich the LHS of equation (\ref{vector_OPE}) between vacuum states, the RHS would be
zero due to Lorentz invariance. For this reason we introduce a small constant background axial field with the quantum numbers  carried by  $J^{5  (3)}$. In the presence of this constant background field, and in the leading order in its value,
we have for the antisymmetric part of the correlator:
\begin{align}\label{two_point_BF}
i \int d^4 x  e^{i q x }  \braket{0|T \left\{ J^{em}_\mu (x) J^{em}_\nu (0) \right\} |0}_{A^{(3)}}=
 i \frac{-N_c}{12 \pi^2}
\left( 1 + C_{Bj}  \frac{2 \pi^2 }{3} ~ \frac{f_\pi^2}{q^2}+ ... \right) \varepsilon_{\mu \nu \rho \sigma} q^\rho A^{(3) \sigma} \,.
\end{align}
Note that the first  term on the RHS is the one loop axial anomaly. The next term is the leading
power correction, and higher power corrections are denoted by the  dots (note that even when
the operator expression for the axial anomaly has an exact one-loop form, the anomalous matrix
element in the three-point function  can receive higher loop perturbative,  as well as non-perturbative corrections
\cite {SVZ_anomaly,Anselm,Mikhailov}). The numerical coefficient of the
$1/q^2$ term is fixed from the axial-axial correlation function in the $\chi$PT
framework (our coefficient differs from that of Ref. \cite{Ioffe:2005ym} by an extra factor of $1/3$):
\bb\label{ChPT_axial}
i \int d^4 x e^{i p x}  \braket{0| T \left\{ J^{(3)}_{5 \mu} (x) J^{(3)}_{5 \nu} (0) \right\} | 0} =
- \frac{N_c}{9} \left( p^2 \eta_{\mu \nu} - p_\mu p_\nu \right) \frac{f_\pi^2}{p^2}\,...
\ee
We are now ready  to proceed to the derivation of these results from holographic QCD.


\vspace{0.2in}

\section{Holographic Crewther-Broadhurst-Kataev Relation}



\subsection{The Gravity Dual}


For simplicity we will only consider two light  quarks and set their masses to zero. Classically this model exhibits global $U_L(2) \otimes U_R(2)$ symmetry and corresponding eight currents are conserved. To describe the correlation functions of these currents in the holographic framework we introduce $4+4$ vector fields,  $L_M$ and $R_M$,
living in the 5D AdS bulk and promote the global $U_L(2) \otimes U_R(2)$ symmetry to  local gauge invariance for these vector fields. It is more convenient to work with their linear combinations $L = V + A$ and $R = V - A$, referred as Vector $(V)$ and Axial $(A)$ fields. These 5D fields will be the actual duals to the 4D vector and axial currents. The  5D AdS space action consists of three pieces, the Yang-Mills, Chern-Simons,  and a boundary term:
\bb
S = S_{YM} + S_{CS} + S_{Boundary} .
\ee
The boundary piece  is introduced to recover the gauge invariance for the vector fields entering the  5D
Chern-Simons term. The relevant part of the action reads:
\bb
S_{YM}^{(2)} \supset - \frac{1}{4 g_5} \int d^4 x \int_{0}^{z_m} dz ~ \frac{1}{z}  \left[ V^a_{MN} V^{a ~ MN} + A^a_{MN} A^{a~MN} \right] ,
\ee
\bb\label{CS_rel}
S_{CS} + S_{Boundary} \supset \frac{3 \kappa}{2 g_5} \varepsilon^{\mu \nu \rho \sigma}
\int d^4 x \left[  \int^{z_m}_0 dz
\left( \partial_z V_\mu ~ \partial_\nu V_\rho^{(3)} - \partial_z V_\rho^{(3)} ~ \partial_\nu V_\mu \right) A^{(3)}_\sigma  \right] .
\ee
Here the capital Latin indices run through $(0,1,2,3,z)$, with $z$ being the bulk coordinate,
while the Greek indices take the values, $(0,1,2,3)$.\footnote{Upper lower case  Latin indices run from 0 to 3, where 0 corresponds to an iso-scalar part and $1,2,3$ to iso-vector ones; 0 is suppressed from iso-scalars.}
The indices  in $S_{YM}^{(2)}$ are contracted via
the 5D flat metric (in our conventions the Minkowski metric is mostly negative,
$\eta_{\mu \nu} = (1,-1,-1,-1)$). The parameter $z_m$ is the position of IR brane, which is introduced in order to model the confining scale. Matching  the results obtained from the 5D gravity to the 4D logarithm in the polarization  function,  and to the 4D  axial  anomaly require: $g_5 = 12 \pi^2 /N_c$ and $\kappa=1$. Furthermore,  we impose
the axial gauge:
\bb
L_z = R_z = 0 . \5\5 \rightarrow \5\5 V_z = A_z = 0 .
\ee
The vector currents are conserved and we only need to consider the transverse part of $V^a_\mu$. It's easier to work in the momentum space:
\bb
V_\mu^a (x,z) =
\frac{1}{(2\pi)^4} \int d^4 q ~ e^{- i q x}  \mathcal{T}_\mu^\alpha (q) V_\alpha^a (q) ~ v(q,z) ,
\ee
\bb
A_\mu^b (x,z)  =
\frac{1}{(2\pi)^4} \int d^4 p ~ e^{- i p x } A_\alpha^b (p)
\Big{(} \mathcal{T}^\alpha_\mu (p) ~ a(p,z) + \mathcal{L}^\alpha_\mu (p) ~ \pi(p,z)
\Big{)} .
\ee
We also split off the sources,  $V_\alpha^a (q),~A_\alpha^a (q)$, from their bulk profiles $v(q,z),a(q,z),\pi(q,z)$.
Here, $\mathcal{T}$ and $\mathcal{L}$ are transverse and longitudinal projectors (see \cite{Colangelo:2011xk} for a similar parametrization) and $a=0,1,2,3; \5 b=1,2,3$.
\bb
\mathcal{T}^\alpha_\mu (p) = \delta^\alpha_\mu - \frac{p^\alpha p_\mu}{p^2}; \5\5\5
\mathcal{L}^\alpha_\mu (p) = \frac{p^\alpha p_\mu}{p^2}.
\ee
Then, the linearized equations of motion and boundary conditions for the vector and
axial bulk profiles take the form (as before, $(Q^2 = - q^2 >0)$):
\begin{align}
\partial_z \left( \frac{1}{z} \partial_z v \right) - \frac{1}{z} ~ Q^2 v  = 0 , &
\5\5\5
v (Q,0) = 1 , \5\5\5 \partial_z v (Q,z_m) = 0 ; \\
\partial_z \left( \frac{1}{z} \partial_z a \right) - \frac{1}{z} ~ Q^2 a  = 0 , &
\5\5\5 a (Q,0) = 1 , \5\5\5  a (Q,z_m) = 0 ; \\
\partial_z \left( \frac{1}{z} \partial_z \pi \right) = 0 , & \5\5\5
\pi(0)=1 , \5\5\5\5 \partial_z \pi(z_m) = 0. \label{longitudinal}
\end{align}
Solutions to these equations can be written in terms of the Bessel functions:
\begin{align}\label{profiles}
v(Q,z) = & Q z \left(K_1(Q z) + I_1(Q z) \frac{ K_0(Q z_m)}{I_0(Q z_m)} \right) , \\
a(Q,z) =  Q z \Big{(} K_1(Q z) & -  I_1(Q z)  \frac{ K_1(Q z_m)}{I_1(Q z_m)} \Big{)} , \5\5\5
\pi(z) = 1 .
\end{align}
Note that the boundary condition on the IR brane for the longitudinal component of the axial profile coincides with the one for a vector. This is a consequence   of  the $z$-component of the equation of motion for the bulk axial field, $\partial_z \partial_\mu A^\mu (x,z)=0$,   that requires the longitudinal  part to be independent of $z$.
In that case, chiral symmetry is broken only spontaneously, and not explicitly,
consistent with massless QCD.

\vspace{0.2in}


\subsection{The Meson Spectra}


One of the successes of holographic QCD is an impressive approximate description of the meson mass spectra \cite{Erlich:2005qh,DaRold:2005mxj,Hirn:2005nr,Domokos:2009cq}. To get it in the axial channel,
we calculate the transverse part of the two point functions using the standard holographic dictionary
\begin{align}\label{two_point_vector_AdS}
\nonumber i \int d^4 x e^{i p x}  \braket{0| T \left\{ J^a_{5 \mu} (x) J^b_{5 \nu} (0) \right\} | 0}^{H}  & =
\frac{1}{g_5} \frac{1}{z} a'(\sqrt{-p^2},z) \delta^{ab} \mathcal{T}_{\mu \nu} (p)
\Big{|}_{z = \epsilon_{UV} \rightarrow 0} = \\
{} & = \left( p^2 \eta_{\mu \nu} - p_{ \mu } p_{ \nu } \right) \frac{- \delta^{a b}}{2 g_5}
 \Pi_5 (p^2) ,
\end{align}
\bb
\Pi_5 (p^2) = 2 \gamma +   \log{\left(\frac{\epsilon_{UV}^2}{4} \right) } - 2
\frac{K_1\left( z_m \sqrt{-p^2} \right)}{I_1\left( z_m \sqrt{-p^2} \right)} +  \log{(-p^2)} ,
\ee
where $\epsilon_{UV}$ is a $UV$ cut-off scale. This expression captures well the correct leading perturbative
log behavior  at high energies, but does not contain power corrections -- all the non-perturbative terms are
exponentials encoded in the Bessel functions.

\begin{figure}[t]
	\centering
	\includegraphics[width=0.5\textwidth]{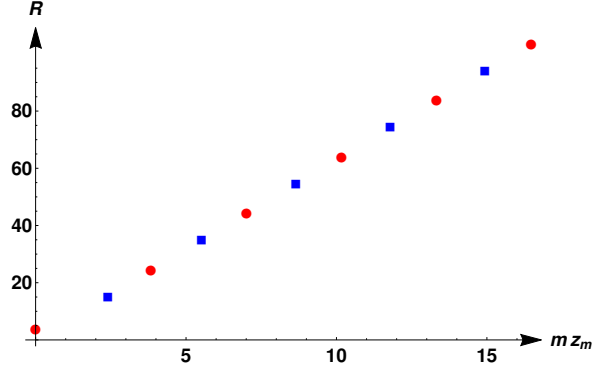}
	\caption{The residues as a function of the poles. The red circles are  for the axial sector, while the
	blue squares for  the vector sector.}\label{residues}
\end{figure}

We should keep in mind that the momentum square
$p^2 <0$,  and to get the particle spectrum analytic continuation to the $p^2>0$ region is needed. Physical particles correspond to the poles of $\Pi_5$ for $p^2 \geq 0$. These are the zeros of the Bessel $J_1$. The masses of axial mesons and residues corresponding to them are ($c_n^i$ is the $n^{th}$ zero of the Bessel $J_i$ ):
\bb
m_n^5 = \frac{c_n^1}{z_m}\, , \5\5\5 R^5_n = - 2 \pi ~ c_n^1 \frac{Y_1 (c_n^1)}{J_0 (c_n^1)}\,.
\ee
The $J_1(x)$ function has its first zero at $x=0$, corresponding to a massless pion with the residue, $R_1^5=4$. The next state should be the $a_1(1260)$ meson.
 The residues are  all positive (see Fig. \ref{residues}). By comparing the residues of the poles in  (\ref{two_point_vector_AdS}) and (\ref{ChPT_axial}) we get the following relation:
\bb\label{duality}
\frac{1}{z_m} = \sqrt{\frac{2}{3}} \pi f_\pi .
\ee
The numerical value $f_\pi = 92.3 MeV$ leads to $z_m^{-1} = 237 MeV$, which is a good
approximation to  $\Lambda_{QCD}$. Moreover, the present setup  enables us to interpret $z_m^{-1}$ as the scale at the chiral symmetry breaking (which is close  to the confinement scale in QCD),  so that $z_m^{-1}$ should be dual to the quark condensate. The numerical value of quark condensate is $\braket{\bar{\psi} \psi} \approx (-240 MeV)^3$ so the predicted value looks  good.

Last but not least, we can  perform a similar analysis to calculate the vector-vector correlator,
with the result for the transverse formfactor:
\bb
\Pi (p^2) = 2 \gamma +   \log{\left(\frac{\epsilon_{UV}^2}{4} \right) } + 2
\frac{K_0\left( z_m \sqrt{-p^2} \right)}{I_0\left( z_m \sqrt{-p^2} \right)} +  \log{(-p^2)} \,,
\ee
and  find the masses and residues of the vector particles:
\bb
m_n = \frac{c_n^0}{z_m}\, , \5\5\5 R_n = 2 \pi ~ c_n^0 \frac{Y_0 (c_n^0)}{J_1 (c_n^0)} \,.
\ee
Since $J_0$ does not have a zero at the origin, there is no massless vector state. This is consistent with QCD.
On the other hand, like in the axial-axial two-point function the above expression captures the  leading perturbative
log behavior  at high energies, but has no power corrections; the non-perturbative terms are
exponentials encoded in the Bessel functions. Thus,  holographic gauge theory differs from QCD
where the power corrections determine the properties of hadrons \cite {SVZ}.

From the above expressions it is possible to predict the ratio of the masses of the vector $\rho$ and  the axial
$a_1$ meson: $\frac{m_\rho}{m_{a_{1}}}= \frac{c_1^0}{c_2^1} \approx 0.63$, which is in agreement with the
observed ratio, ${m_{\rho (770)}\over m_{a_1(1260)} }= {775\,MeV\over 1230\,MeV}\simeq
0.63$.\footnote{By $\rho$ meson we mean the first pole that appears in the vector-vector two point function
corresponding to
$\rho^0, \rho^\pm$ and $\omega$, all degenerate here since  we ignored the electromagnetism, the
up and down quark masses, and the strange quark.}
Even though this model correctly predicts the ratio of the $\rho$ and $a_1$ meson masses,
the  predicted absolute values of these quantities,  $m_\rho = 570 MeV$ and $m_{a_1} = 910 MeV$,  are off
by 25\% or so.  This is perhaps expected   since the holographic model is minimalistic, has zero quark masses,
and does not account for the gluon condensate; the latter makes a contribution to the vector meson
masses in the QCD sum rules \cite {SVZ}.   Accounting for these effects might potentially bring the masses of the
mesons closer to the observed ones, as suggested  by the relations derived in \cite{Krasnikov:1982ea}.

\vspace{0.2in}


\subsection{Non-perturbative Physics in CBK}

\begin{figure}[H]
	\centering
	\includegraphics[width=0.5\textwidth]{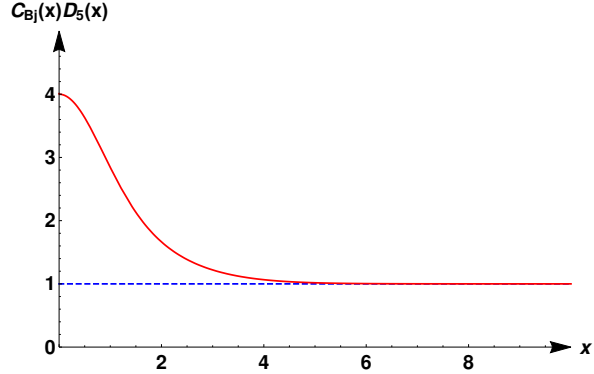}
	\caption{The product of the Bjorken and Adler functions,  as a function of the dimensionless momentum
	square, ($x= z_m \sqrt{-q^2}$). The numerical value at $x=0$ comes from the residue of a
	massless pion.}\label{FIG_CBK}
\end{figure}

In order to calculate the Adler and Bjorken functions we need the symmetric and antisymmetric parts of the axial and electromagnetic current two point functions,  respectively.  We already got the symmetric part for the axial current correlator in (\ref{two_point_vector_AdS}). Comparing this formula with (\ref{axial_OPE}), and using
$D_5 = d \Pi_5 /d \ln (P^2)$, we get:
\bb\label{adler}
D_5 (p^2) = 1 + \frac{1}{I_1\left( z_m \sqrt{-p^2} \right)^2} \,.
\ee
Finding the exact Bjorken function is more tricky, one needs to calculate the electromagnetic current two point function in the presence of a constant background axial field. The relevant part of the 5D action is the Chern-Simons term. Although the background field is constant on the UV boundary, one  still has to consider its non-constant bulk profile. A bit of a
calculus yields:
\begin{align}\label{two_point_BF_AdS}
\nonumber {} & i \int d^4 x e^{i q x} \braket{0| T \left\{ J^{em}_\mu (x) J^{em}_\nu (0) \right\} | 0}_{A^{(3)}}^{H} = \\
{} = & \frac{i \kappa}{g_5}
\int^{z_m}_0 dz   ~
\Big{(} \frac{3}{4} a(0,z) + \frac{1}{4} \pi(z) \Big{)} ~ \partial_z v^2 (q,z)  ~
\varepsilon^{\mu \nu \rho \sigma} q_{\rho} A_\sigma^{(3)} =  \\
\nonumber = & - \frac{ i \kappa }{ g_5}
\left[ 1 + \frac{1}{ q^2 z_m^2} - \left( \frac12 + \frac{1}{ q^2 z_m^2} \right) I_0 \left( z_m \sqrt{-q^2} \right)^{-2} \right]
\varepsilon^{\mu \nu \rho \sigma} q_{\rho} A_\sigma^{(3)} .
\end{align}
One has to compare this formula with its  field theory
counterpart (\ref{two_point_BF}), which together with (\ref{duality}) gives the exact Bjorken function:
\bb\label{bjorken}
C_{Bj} (q^2) =  1
-  \left( 1 + \frac{q^2 z_m^2}{ 2} \right) I_0 \left( z_m \sqrt{-q^2} \right)^{-2} .
\ee
Multiplying (\ref{adler}) by (\ref{bjorken}) gives the holographic CBK relation:
\begin{align}\label{CBK_H_1}
 {} & C_{Bj} (q^2) ~ D_5 (q^2) =  1
-  \left( 1 + \frac{q^2 z_m^2}{ 2} \right) I_0 \left( z_m \sqrt{-q^2} \right)^{-2} + \\
\nonumber {} & +  I_1 \left( z_m \sqrt{-q^2} \right)^{-2}
-  \left( 1 + \frac{q^2 z_m^2}{ 2} \right) I_0 \left( z_m \sqrt{-q^2} \right)^{-2} I_1 \left( z_m \sqrt{-q^2} \right)^{-2} .
\end{align}
This expression is exact! The two asymptotic values of this relation are (See Fig. \ref{FIG_CBK}):
\bb
C_{Bj} (q^2) ~ D_5 (q^2) \Big{|}_{q^2 \rightarrow \infty} =1; \5\5\5
C_{Bj} (q^2) ~ D_5 (q^2) \Big{|}_{q^2 \rightarrow 0} = 4\, .
\ee
The hight momentum limit corresponds to perturbative QCD; hence, the obtained value of the product is consistent with the pQCD calculations in the conformal limit  \cite{Crewther:1972kn,Broadhurst:1993ru,Gabadadze:1995ei,Brodsky:1995tb,Kataev:1996ce}. The low momentum-square formula is one of our main results;
we emphasize that the low energy numerical value is  determined by the residue of a
massless pion. At intermediate momenta, $q^2 \sim z_m^{-2}$,
the product $C_{Bj} (q^2) ~ D_5 (q^2)$   deviates from 1 by terms related to  various resonances. The
latter suggest that  in QCD  the CBK relation should be  modified  by  non-perturbative corrections, in similarity with
eq. (\ref {CBK_H_1})

\vspace{0.2in}


\subsection{Consistency with Three-point Function}


The 5D Chern-Simons term can be used to calculate the anomalous three point function as well. A lengthy but straightforward calculation gives (see Appendix \ref{app_A} for the details):
\begin{align}\label{three_point_H}
{} & i \int d^4 x d^4 y e^{i k_1 x + i k_2 y}
\braket{0| T \left\{ J^{em}_\mu (x) J^{em}_\nu (y) J^{5  (3)}_\rho (0) \right\} | 0}^H = \\
\nonumber  = & F_3^{H} ~ \varepsilon_{\mu \nu \rho \tau} q^\tau + F_2^{H} \left( p_\mu \varepsilon_{\nu \rho \sigma \tau} - p_\nu \varepsilon_{\mu \rho \sigma \tau} \right) p^\sigma q^\tau
+ F_1^{H} \left( q_\mu \varepsilon_{\nu \rho \sigma \tau} + q_\nu \varepsilon_{\mu \rho \sigma \tau} \right) p^\sigma q^\tau ,
\end{align}
where:
\bb\label{F_1_AdS_def}
F_1^{H} =  \frac{\kappa }{g_5} \frac{ R_T (p^2,q^2) }{p^2 + q^2} ,
\ee
\bb\label{F_2_AdS_def}
F_2^{H} = - \frac{\kappa }{g_5} \left(\frac{q^2}{p^2} \frac{R_T (p^2,q^2)}{p^2 + q^2} - \frac{R_L (p^2,q^2)}{p^2} \right) ,
\ee
\bb\label{F_3_AdS_def}
F_3^{H} = - \frac{\kappa }{g_5} R_L (p^2,q^2) ;
\ee
\bb\label{transverse_FF}
R_T (p^2,q^2) \equiv  \int^{z_m}_0 dz ~ a(2 \sqrt{-p^2},z) ~ \partial_z v^2 (\sqrt{-p^2-q^2},z) ,
\ee
\bb\label{longitudinal_FF}
R_L (p^2,q^2) \equiv  \int^{z_m}_0 dz ~ \pi (z) ~ \partial_z v^2 (\sqrt{-p^2-q^2},z) .
\ee
The integrand in (\ref{transverse_FF}) is a product of three Bessel functions with different arguments, for which  it's
hard find a general analytic  expression. Luckily, it's sufficient  for us to have the integral evaluated   at $q^2 \gg p^2$. For the details of the calculation see Appendix \ref{app_B}.  Having this done, we can then
follow the procedure used in Section 2, and get:
\bb\label{CBK_H_2}
C_{Bj} (Q^2) ~ D_5 (4 P^2) \Big{|}_{q^2 \rightarrow + \infty}  = 1 +
\frac{1}{I_1 \left( 2 z_m \sqrt{-p^2} \right)^2} \,.
\ee
Dividing this formula by (\ref{adler}), we get the asymptotic value of the Bjorken function in agreement
with the expression (\ref{bjorken}).

\vspace {0.2in}


\subsection{A Low Energy Relation}


Let us look at the antisymmetric part of the  two point function  of the electromagnetic fields. As before, we
assume the presence of a constant background axial field with the quantum numbers of the pion current.
In  QCD  so defined, and in the leading order in the external field,  we can split this correlator as
a sum of the axial anomaly  plus the rest:
\begin{align}\label{gen_axial_two_point}
i   \int d^4 x  e^{i q x }  \braket{0|T \left\{ J^{em}_\mu (x) J^{em}_\nu (0) \right\} |0}_{A^{(3)}}=  - i \frac{N_c}{12 \pi^2}
\left[ 1 + f (q^2)  \right] \varepsilon_{\mu \nu \rho \sigma} q^\rho A^{(3) \sigma} ,
\end{align}
where the first term is due to the  axial anomaly,  while  the  form-factor $f$  parametrizes
the rest, including all the non-perturbative parts. This two-point function  can be calculated in homographic QCD
and is given  by the first equation in  (\ref{two_point_BF_AdS}).  Comparison  of (\ref {gen_axial_two_point})
with (\ref{two_point_BF_AdS}) gives:
\bb\label{gen_axial_two_point_pol}
1  + f (q^2) =
1 + \frac{1}{ q^2 z_m^2} - \left( \frac12 + \frac{1}{ q^2 z_m^2} \right) I_0 \left( z_m \sqrt{-q^2} \right)^{-2} .
\ee
In the limit $q^2 \rightarrow 0 $, the RHS of (\ref{gen_axial_two_point_pol}) vanishes and we are left with:
\bb\label{sum_rule}
1   =-  f (q^2) \Big{|}_{q^2=0}\, .
\ee
The latter is a low energy relation that  connects the axial anomaly, on the LHS,  with the rest of the contributions
to the two-point function in  holographic QCD, represented on the RHS. While the high-$q^2$ behavior of $f(q^2)$
can be calculated in QCD, the $q^2\to 0$ limit is  perturbatively  unaccessible.

\vspace{0.2in}

\acknowledgments
We'd like to thank Sophia Domokos and Andrei Kataev for  stimulating and helpful
correspondence, and Arkady Vainshtein for useful discussions and suggestions.
The work was supported in part by NSF grant PHY-1620039.


\appendix

\section{Appendix: Conformal Three Point Functions}\label{app_A}

Adopting the kinematical condition  that  the square momenta of the two photons equal to each other,
we can write the most general expression  for the anomalous three point function as follows
\cite{Gabadadze:1994tn}:
\begin{align}\label{three_point}
\Delta^{em,em,(3)}_{\mu \nu \rho} = & i \int d^4 x d^4 y e^{i k_1 x + i k_2 y}
\braket{0| T \left\{ J^{em}_\mu (x) J^{em}_\nu (y) J^{5  (3)}_\rho (0) \right\} | 0} = \\
\nonumber  = & F_3 ~ \varepsilon_{\mu \nu \rho \tau} q^\tau + F_2 \left( p_\mu \varepsilon_{\nu \rho \sigma \tau} - p_\nu \varepsilon_{\mu \rho \sigma \tau} \right) p^\sigma q^\tau
+ F_1 \left( q_\mu \varepsilon_{\nu \rho \sigma \tau} + q_\nu \varepsilon_{\mu \rho \sigma \tau} \right) p^\sigma q^\tau \,,
\end{align}
were $k_1 = p+q$ and $k_2 = p-q$ and the kinematic constraint specified above translates into,
$p \cdot q  =0$. As before, the quarks are massless. One loop expressions for the form factors $F_i$ can be found in \cite{Rosenberg:1962pp}. $F_1$ and $F_2$ are finite functions, while $F_3$ is defined through the vector Ward identity and is just a number
\bb
F_1^{PT}  =  \frac{N_c}{6 \pi^2} \int_0^1 dx  \int_0^{1-x} dy \frac{(x+y) -(x+y)^2}{- (x+y) \left( p^2 + q^2 \right) + p^2 \left(x-y\right)^2 + q^2 \left(x+y\right)^2 } ,
\ee
\bb\label{F_2_PT_def}
F_2^{PT}  =  \frac{N_c}{6 \pi^2} \int_0^1 dx  \int_0^{1-x} dy \frac{x+y-(x-y)^2}{- (x+y) \left( p^2 + q^2 \right) + p^2 \left(x-y\right)^2 + q^2 \left(x+y\right)^2 } ,
\ee
\bb\label{anomaly}
F_3^{PT} = - \left( p^2 F_2^{PT} + q^2 F_1^{PT} \right) =  \frac{N_c}{12 \pi^2} .
\ee
Within the holographic framework, conformal limit corresponds to $z_m \rightarrow + \infty$, in this case we recover the exact  AdS metric, and the results of the holographic calculation  for the three-point function
should agree with the one-loop perturbative field theory calculation, since the latter is all-loop
exact in a conformal theory \cite {Schreier:1971um}. Integrating the profile functions (\ref{transverse_FF}) and (\ref{longitudinal_FF}) we calculate:
\bb\label{transverse_FF_C}
R_T^{AdS} (p^2,q^2) = - 1 + \frac{\sqrt{\pi}}{2} G_{3,3}^{3,2}\left(
\begin{array}{c}
	0,0,\frac{3}{2} \\
	0,1,2 \\
\end{array}
\Big{|} 1 + \frac{q^2}{p^2} \right) ,
\ee
\bb\label{longitudinal_FF_C}
R_L^{AdS} (p^2,q^2) = - 1 \,,
\ee
where $G$ stands for the Meijer G-function. After plugging these  expressions
into (\ref{F_1_AdS_def})-(\ref{F_3_AdS_def}),  we obtain:
\bb\label{anomaly_AdS}
F_3^{AdS} =  \frac{N_c}{12 \pi^2} ,
\ee
\bb
F_2^{AdS}  =
- \frac{N_c}{12 \pi^2} \frac{1}{p^2 + q^2}
\left[ 1 + \frac{\sqrt{\pi}}{2} \frac{q^2}{p^2}
G_{3,3}^{3,2}\left(
\begin{array}{c}
	0,0,\frac{3}{2} \\
	0,1,2 \\
\end{array}
\Big{|} 1 + \frac{q^2}{p^2} \right)
\right] .
\ee
Furthermore, by doing a numerical analysis of these expressions as well as
perturbative expansions,  we convinced ourselves that the relation
\bb
F_2^{PT} = F_2^{AdS} \equiv F_2\,,
\ee
holds for arbitrary  $p$ and $q$; hence,  the AdS/CFT conjecture works well in this case!
It is possible to express $F_2$ in terms of the DiLogarithm  as follows:
\bb\label{F_2_PT_FULL}
F_2 =  - \frac{N_c}{12 \pi^2} \frac{1}{q^2} \left\{ \frac{1}{2 x} - \frac{1+3x}{4x} \ln \left(\frac{4x}{1+x}\right)
- \frac{\left( 1+x \right) \left( 1 - 3 x \right)}{4 ~x^{3/2}}
\text{Im} \left[ \text{Li}_2 \left( e^{2 i \tan^{-1} \sqrt{x}} \right) \right] \right\} ;
\ee
where $ x \equiv p^2/q^2 $ and $``\text{Im}"$ stands for the imaginary part.

\vspace{0.2in}


\section{Appendix: Beyond Conformal Limit}\label{app_B}

\begin{figure}[H]
	\centering
	\begin{subfigure}[b]{0.4\textwidth}
		\includegraphics[width=\textwidth]{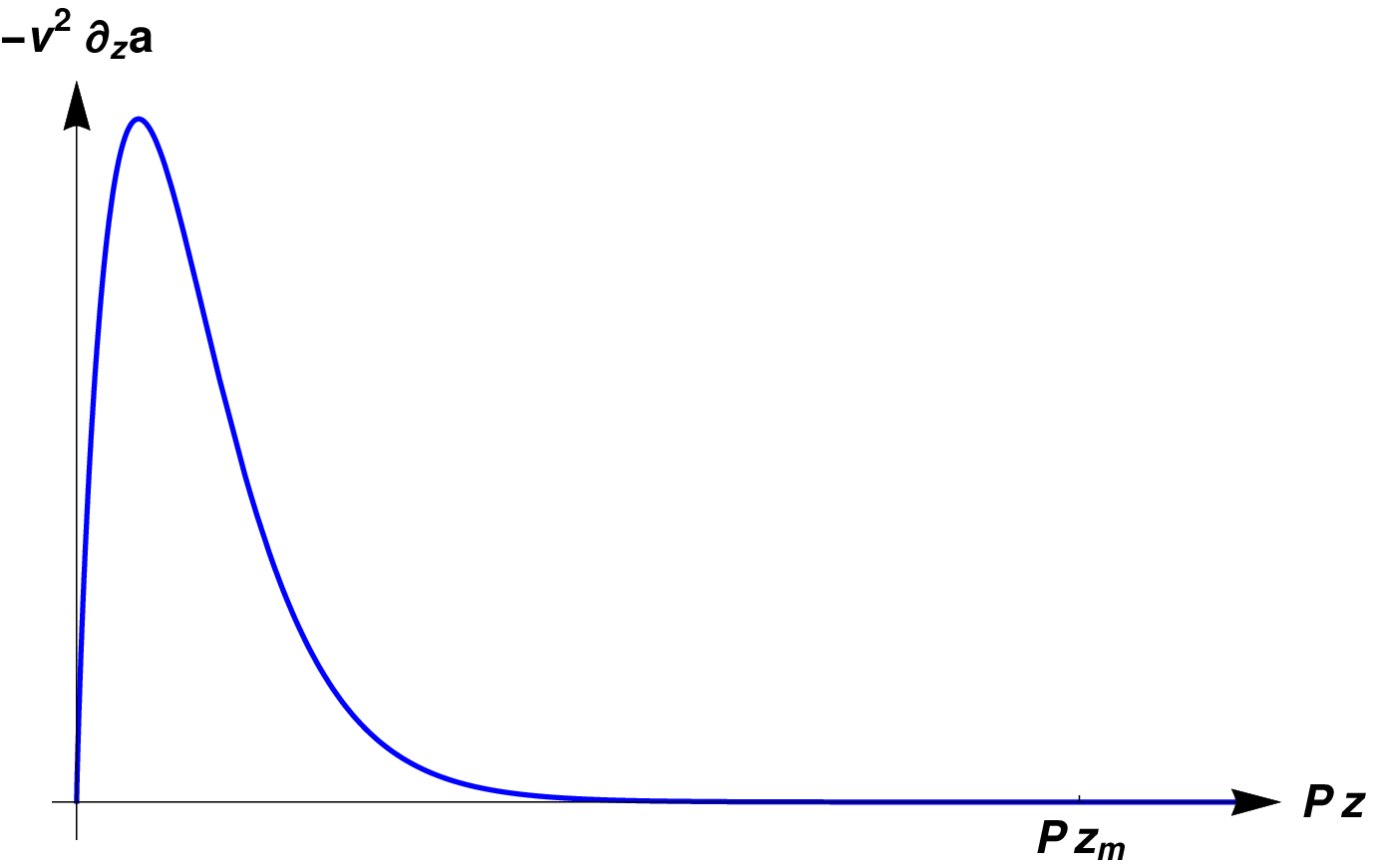}
		\caption{}
	\end{subfigure}
	\begin{subfigure}[b]{0.4\textwidth}
		\includegraphics[width=\textwidth]{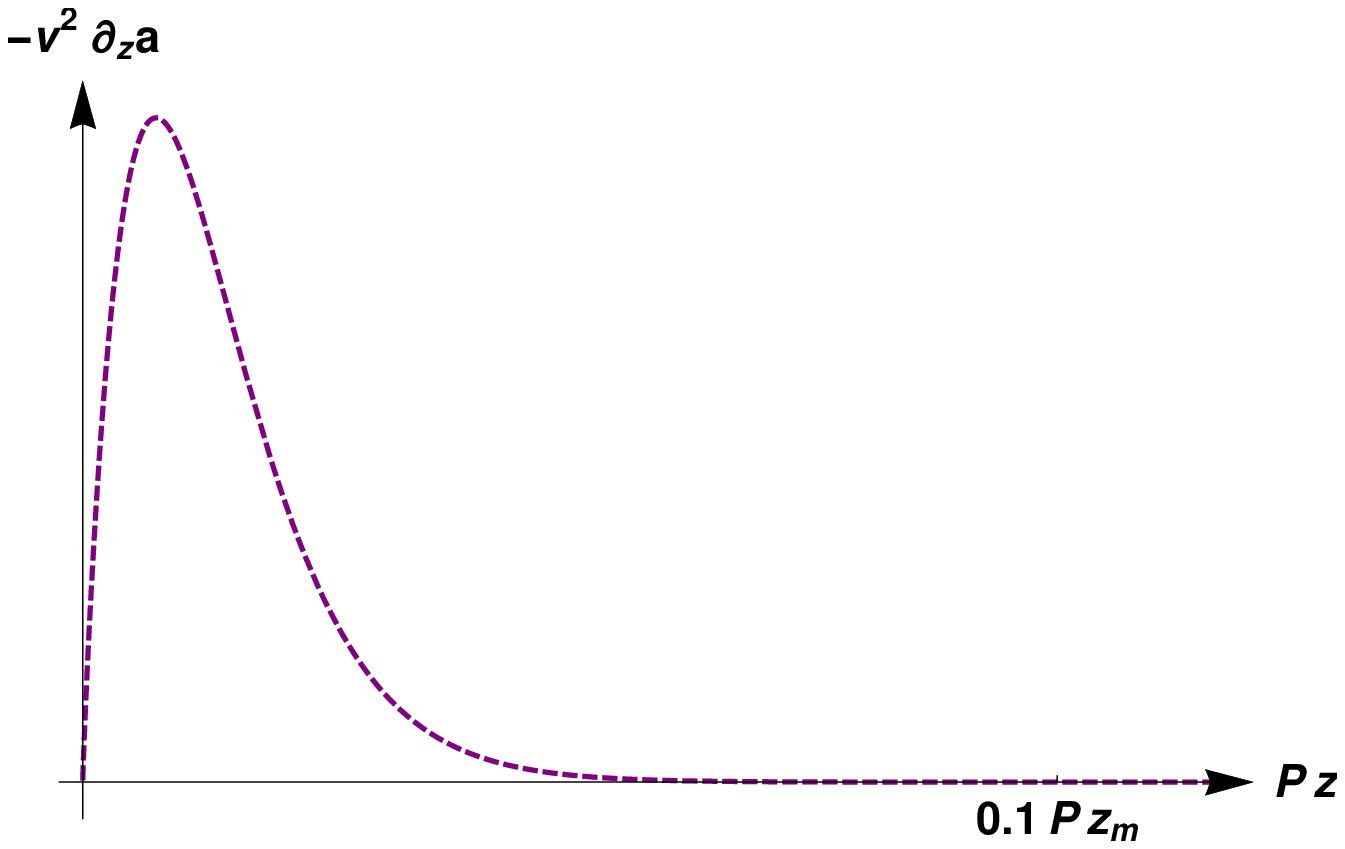}
		\caption{}
	\end{subfigure}
	\caption{The plots of the integrand in (\ref{trans_app}). In the left solid blue plot (a) corresponds to $q^2/p^2=100$ and the right dashed purple (b) to $q^2/p^2=10000$. In both cases $-p^2 z_m^2 =1$. Increasing the value of $-p^2$, while keeping the ratio $q^2/p^2$ fixed, makes our approximation even better. }\label{ints}
\end{figure}

Calculation   of the  integral in (\ref{transverse_FF}) is not an easy task, since it contains a
product of three Bessel functions with different arguments. To tackle this problem, let us  first perform
partial integration and write:
\bb\label{trans_app}
R_T (p^2,q^2) = -1 -  \int^{z_m}_0 dz ~ v^2 (\sqrt{-p^2-q^2},z) ~ \partial_z a(2 \sqrt{-p^2},z) .
\ee
In the approximation  when $-q^2 \gg z_m^{-2}$, or $-p^2 \gg z_m^{-2}$, the integrand gets localized near the UV boundary and the integral becomes essentially independent of the IR  cut-off $z_m$ (See Fig. \ref{ints}). Using this fact, we extend the range of integration from $[0,z_m]$ to $[0,+\infty)$. Note that in the range $[z_m,+\infty)$ the integrand has bad behavior and the integral is divergent. \textit{Mathematica} can explicitly solve this integral. After removing the divergent part, we find non-conformal corrections to (\ref{transverse_FF_C}) and (\ref{longitudinal_FF_C}):
\begin{align}
R_T (p^2,q^2) = & - 1 + \frac{\sqrt{\pi}}{2} G_{3,3}^{3,2}\left(
\begin{array}{c}
0,0,\frac{3}{2} \\
0,1,2 \\
\end{array}
\Big{|} 1 + \frac{1}{x} \right) +  \\
\nonumber {} & +  \left(1+x\right)
\left(1+3x +\frac{\left(1+x\right) \left(-1+3x\right)}{\sqrt{x}} \tan ^{-1} (\sqrt{x} )\right)\frac{K_1(2z_m \sqrt{-p^2})}{I_1(2z_m \sqrt{-p^2})} + \\
\nonumber {} & + \mathcal{O} \left( e^{ - 2 z_m \sqrt{ -p^2 - q^2} } \right) , \\
R_L (p^2,q^2) = & -1 + \frac{1}{I_0 \left(z_m \sqrt{-p^2-q^2} \right)^2 } =- 1 + \mathcal{O} \left( e^{ - 2 z_m \sqrt{ -p^2 - q^2} } \right) .
\end{align}
Here again $ x \equiv p^2/q^2 $. The small terms come from the ratios of the Bessel functions;
since we chose $-q^2 \gg z_m^{-2}$, or $-p^2 \gg z_m^{-2}$,  the  arguments of the Bessel functions are
large,  and the functions themselves are exponentially suppressed.  After plugging these into (\ref{F_1_AdS_def}-\ref{F_3_AdS_def}),  and assuming that  $p^2/q^2 \ll 1$, we get:
\begin{align}\label{F_2_H}
 F_2^{H}  \approx & \frac{N_c}{9 \pi^2} \frac{1}{q^2} \left[ -\frac{11}{12} + \ln{\left( \frac{4 p^2}{q^2} \right)} - 2 \frac{K_1 \left( 2 z_m \sqrt{-p^2} \right)}{I_1 \left( 2 z_m \sqrt{-p^2} \right)}  \right. + \\
\nonumber {} & \left. +\frac{2}{5} \frac{p^2}{q^2} \left( -\frac{503}{120} + \ln{\left( \frac{4 p^2}{q^2} \right)} - 2 \frac{K_1 \left( 2 z_m \sqrt{-p^2} \right)}{I_1 \left( 2 z_m \sqrt{-p^2} \right)}  \right) \right] +\mathcal{O} \left( \frac{p^4}{q^6} \right) ,
\end{align}
\bb
F_3^{H} \approx   \frac{N_c}{12 \pi^2}  + \mathcal{O} \left( e^{ - 2 z_m \sqrt{ -p^2 - q^2} } \right) .
\ee
The first two terms in each line in the expansion of $F_2^{H}$ come from the conformal part,  the rest
are non-perturbative corrections. Note that the non-perturbative effects also renormalize the anomalous part.
Taking logarithmic derivative of (\ref{F_2_H}) gives (\ref{CBK_H_2}).



\end{document}